\documentclass[a4paper,12pt]{article}

\usepackage{amssymb, amsmath, amsthm, algorithm,subfig}
\usepackage{multirow}
\usepackage{verbatim, url, amssymb, varioref,  calc, listings, amsmath, color}
\definecolor{hotpink}{rgb}{0.9,0,0.5}
\usepackage[colorlinks,urlcolor=blue,citecolor=hotpink,linkcolor=blue]{hyperref}
\usepackage{pgf}
\usepackage{tikz}
\def \bigO{\mathcal{O}}

\def \bigO{\mathcal{O}}

\def \eixx{EI_{xx}}
\def \eiyy{EI_{yy}}
\def \eixxbyl1{\frac{\eixx}{\ell}}
\def \eixxbyl2{\frac{6\eixx}{\ell^2}}
\def \eixxbyl3{\frac{12\eixx}{\ell^3}}
\def \eiyybyl1{\frac{\eiyy}{\ell}}
\def \eiyybyl2{\frac{\eiyy}{\ell^2}}
\def \eiyybyl3{\frac{\eiyy}{\ell^3}}
\def\e{eigenvalue}

\def\l{\lambda}

\usepackage{verbatim}
\usepackage{tkz-fct}
\usetikzlibrary{arrows,automata}
\title{A clustering tool for interrogating finite element models based on eigenvectors of graph adjacency}
\date{}
\author{Ramaseshan Kannan\footnote{ramaseshan.kannan@arup.com} \\Algorithms and Numerical Analysis\\ Digital Technology,  Arup}

\mathcode`@="8000 
{\catcode`\@=\active\gdef@{\mkern1mu}}

\def \bigO{\mathcal{O}}

\def\R{\mathbb{R}}

\usepackage{float}
\newfloat{Algorithm}{H}{lop}
\newfloat{mylistings}{H}{myl}
\def\half{\frac{1}{2}}

\newcounter{mylineno}
\makeatletter
\let\oldtabcr\@tabcr

\def\mynewline{\refstepcounter{mylineno}%
                \llap{\footnotesize\arabic{mylineno}\hspace{5pt}}%
               }

\gdef\@tabcr{\@stopline \@ifstar{\penalty%
            \@M \@xtabcr}\@xtabcr\mynewline}

\makeatother

\begin{document}

\maketitle

This note introduces an unsupervised learning algorithm to debug errors in finite element (FE) simulation models and details how it was \textit{productionised}. The algorithm clusters degrees of freedom in the FE model using numerical properties of the adjacency of its stiffness matrix. The algorithm has been deployed as a tool called `Model Stability Analysis' tool within the commercial structural FE suite Oasys GSA\footnote{\url{www.oasys-software.com/gsa}}. It has been used successfully by end-users for debugging real world (FE) models and we present examples of the tool in action.

\section{Introduction}
Structural FE simulation models can often contain errors from a variety of sources.
Some errors are inherent to the process of modelling system behaviour as partial differential equations. Others arise from discretization or are modelling oversights. 
The presence of these errors can invalidate the simulation and impact the safety of structural designs. 

If is therefore imperative to establish the sanity of simulation models. 
There are two reasons why doing so is of growing importance.
Firstly, models are growing rapidly in size and complexity to take advantage of increasing availability of compute power. 
Secondly, they are increasingly being generated using automated processes and hence there is reduced human oversight in how they are put together.

Errors in FE models can manifest either silently or overtly. 
An overt error stops the analysis from proceeding. 
Finding and fixing these errors is time consuming and the practitioner must usually rely on ad-hoc, rule-of-thumb tricks and approaches learnt from experience.
Errors that manifest silently are however more pernicious as their presence is hard to detect and it is harder still to find their cause.  
  

Model Stability Analysis is a tool that uses numerical linear algebra, unsupervised clustering, and rounding error analysis to identify parts of the model that have errors. 
In the next section, we briefly explain the underlying algorithm in a graph theoretic context. 
For a numerical analysis explanation, including proofs of why it works and perturbation analysis, the reader is referred to \cite{kahehiti14}. 

\section{The algorithm}
There are two ansatzes that lead us to identifying the source and cause aforementioned errors.
\begin{enumerate}
	\item The class of errors we are interested in cause \href{https://nhigham.com/2020/03/19/what-is-a-condition-number/}{ill conditioning} of the stiffness matrix. Ill conditioning is the heightened sensitivity of a problem to small changes in the input data. 
	An ill conditioned stiffness matrix consists of entries that are either disproportionately large or small. Sometimes both. Large condition numbers can lead to inaccurate answers when inverting matrices as FE computations are done in double precision \href{https://nhigham.com/2020/05/04/what-is-floating-point-arithmetic/}{floating point arithmetic}.
	\item The eigenvectors of the adjacency matrix of the degree-of-freedom graph contain information that reveals parts of model responsible for the ill-conditioning, and subsequently the modelling errors. 
\end{enumerate}

Therefore we first detect the presence of ill-conditioning using a condition number estimator. 
In technical terms, this step allows us to assess the numerical rank of the matrix.
Next comes the interesting part of determining elements whose definitions cause ill conditioning, which uses an unsupervised learning approach by clustering the degrees of freedom (dofs) informed by eigenvectors of the adjacency matrix. 

Spectral clustering and the use of eigenvectors is a well-known analysis technique for graphs when modelling relationships in areas as disparate as social networks, computational physics, biology or finite element analysis.
Good introductions are \href{https://arxiv.org/abs/0711.0189}{`A Tutorial on Spectral Clustering'} and \cite{von07}. 

The method involves forming the Laplacian or the adjacency\cite{luwisl14} of the graph, and computing the eigenvectors corresponding to certain eigenvalues -- often the first few. 
The eigenvectors have a certain sparsity structure that reveals the clusters in the graph nodes, which in turn correspond to the weakly connected components of the graph. 

\begin{figure} 
	\centering
	\includegraphics[width=0.75\textwidth]{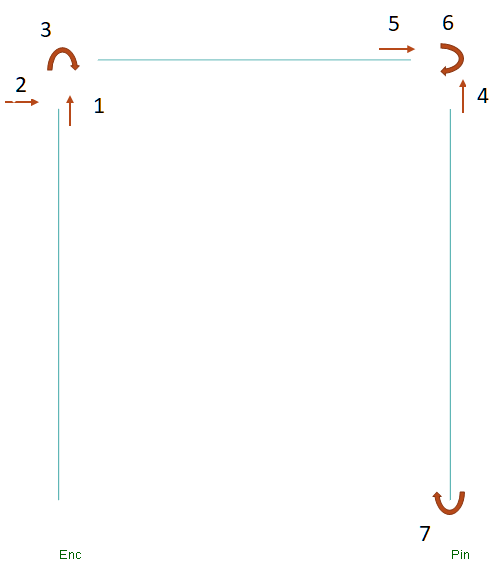} 
	\caption{A frame model with numbered stiffness directions}
	\label{fig-portal-frame}
\end{figure}

For Model Stability Analysis we are interested in identifying elements that share dofs that have either disproportionately small or large stiffness. 
To this end, we developed a variant of spectral clustering that maximizes an energy function across all dofs for each eigenvector of interest.
This is best explained with an example. 
The frame in Figure \ref{fig-portal-frame} is a simple beam model in two dimensions showing dofs corresponding to each stiffness direction.

Assuming a stiffness of $\bigO(k)$ along all directions, the graph of the dof connectivity is Figure \ref{fig-graph}.
The problem we are interested in is, \emph{given a dof with insufficient stiffness, how do we isolate elements to which it is connected?}
If the weights of edges to dof 7 drop to a value $\epsilon \ll k$, the vertex becomes a weakly connected component of the graph, as shown in figure \ref{fig-graph-weaklyconnected}.
When $\epsilon$ becomes $0$ we have disconnected graphs and the corresponding structural model has rigid body modes.

\begin{figure}
\begin{tikzpicture}[<->,>=stealth',shorten >=1pt,auto,node distance=3.8cm,
semithick]
\tikzstyle{every state}=[fill=red,draw=none,text=white]

\node[state] 		 (6)                    {$6$};
\node[state]         (2) [below right of=6] {$2$};
\node[state]         (3) [above right of=6] {$3$};
\node[state]         (4) [left of=6] {$4$};
\node[state]         (5) [right of=6]       {$5$};
\node[state]         (1) [above of=4]       {$1$};
\node[state]         (7) [below left of=6]       {$7$};

\path (6) edge              node {k} (4);
\path (6) edge              node {k} (7);
\path (6) edge              node {k} (2);
\path (6) edge				node {k} (3);
\path (6) edge				node {k} (5);

\path (2) edge [bend right=90]             node {k} (3);
\path (3) edge              node {k} (5);
\path (5) edge              node {k} (2);
\path (1) edge				node {k} (4);
\path (1) edge				node {k} (3);
\path (4) edge				node {k} (7);
\end{tikzpicture}
\caption{DoF connectivity graph for frame in Fig \ref{fig-portal-frame} with edge weights $\bigO(k)$ from stiffness at each dof.}
\label{fig-graph}
\end{figure}
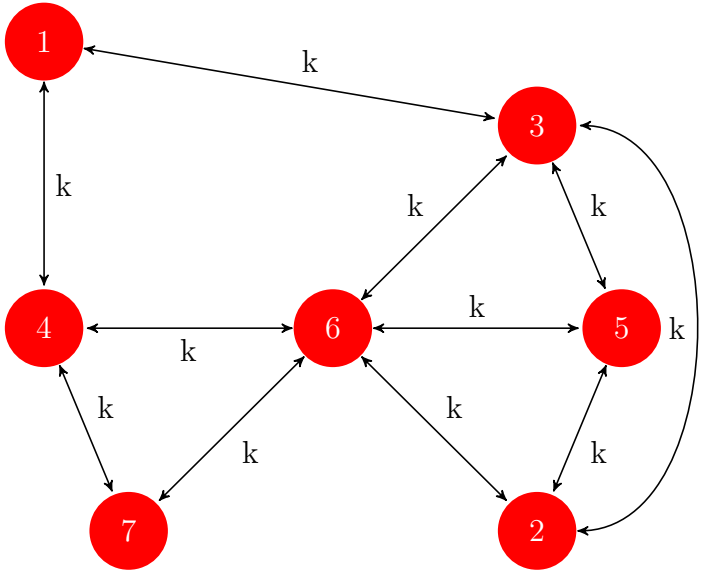
\begin{figure}
	\begin{tikzpicture}[<->,>=stealth',shorten >=1pt,auto,node distance=3.8cm,
	semithick]
	\tikzstyle{every state}=[fill=red,draw=none,text=white]
	
	\node[state] 		 (6)                    {$6$};
	\node[state]         (2) [below right of=6] {$2$};
	\node[state]         (3) [above right of=6] {$3$};
	\node[state]         (4) [left of=6] {$4$};
	\node[state]         (5) [right of=6]       {$5$};
	\node[state]         (1) [above of=4]       {$1$};
	\node[state]         (7) [below left of=6]       {$7$};
	
	\path (6) edge              node {k} (4);
	\path (6) edge              node {k} (2);
	\path (6) edge				node {k} (3);
	\path (6) edge				node {k} (5);
	
	\path (2) edge [bend right=90]             node {k} (3);
	\path (3) edge              node {k} (5);
	\path (5) edge              node {k} (2);
	\path (1) edge				node {k} (4);
	\path (1) edge				node {k} (3);
\begin{scope}[every edge/.style={draw=black,dashed}]
	\path (4) edge				node {$\epsilon$} (7);
	\path (6) edge              node {$\epsilon$} (7);
\end{scope}
	
\begin{scope}[-,node distance = 2.5cm, every edge/.style={draw=blue, line width=2,dashed}]
	\node(99)[left of=7]{};
\node(100)[right of=7]{};

\path (99) edge [bend left=55] node{} (100);

\end{scope}
	\end{tikzpicture}
	\caption{Modified connectivity graph with a weakly connected component. Dashed blue line indicates the clustering boundary we seek.}
	\label{fig-graph-weaklyconnected}
\end{figure}
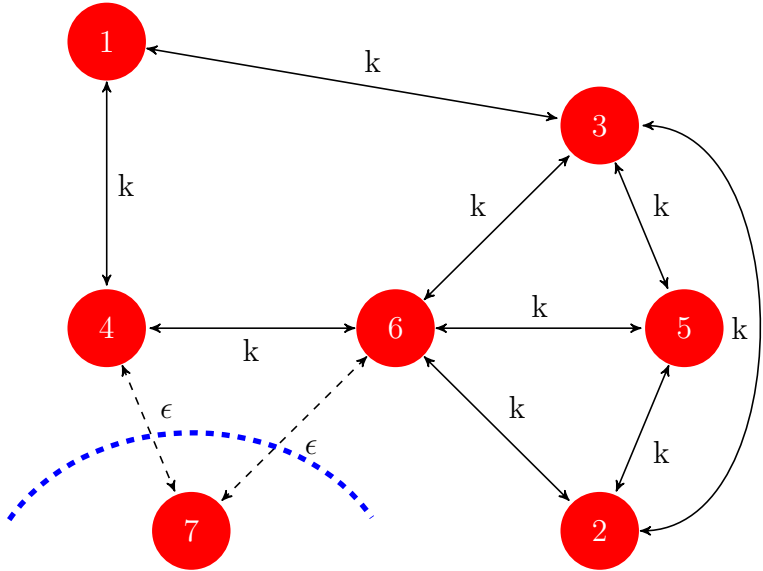
The key to identifying weakly connected/disconnected components is by recognizing that dofs in these components form clusters. 
Normalized eigenvectors corresponding to the null-space have a natural sparsity structure where dofs in clusters have large nonzero values and all other components of the eigenvectors are nearly $0$; therefore we can partition them into two clusters per eigenvector. 
Using perturbation theory we prove in \cite{kahehiti14} that ill-conditioned FE stiffness matrices will have associated adjacency graphs with weakly connected components, and that the corresponding eigenvectors have a required sparsity structure.  

Combining these ideas we derive Algorithm \ref{alg-modelstab}.
\begin{algorithm}
	\caption{Algorithm for model stability analysis.}
	Given a graph of an FE assembly with weakly connected components and a user-specified number of eigenpair, this algorithm returns energy functions $v^{(e)}$ and $s^{(e)}$ for each eigenpair, and their clusters. It uses the following inputs and user-defined parameters:
	\begin{itemize}
		\item Weighted adjacency (structural stiffness) matrix $A$. 
		\item Domain $\Omega$ of all mesh elements.
		\item Transformation+mapping matrix $T$ from element local to global dofs and mapping \(m_{e}\) of \(n_e\) variables that map the locally numbered dofs of element $e$ to the global dof numbering.
		\item $n_s \geq 0$: the number of smallest eigenpairs;
		\item $n_\ell \geq 0$: the number of largest eigenpairs;
		\item $\mathrm{gf} \geq 1$: the spectral gap between a cluster
		of smallest eigenvalues and the next largest eigenvalue.
	\end{itemize}
	\begin{enumerate}
		
		\item\label{alg-calc-r}
		Solve $A u = \lambda u$ using a sparse symmetric eigensolver to find the $n_s$ smallest \e s $\l_1$, $\l_2$, $\dots$, $\l_{n_s}$ and
		$n_\ell$ largest \e s $\l_{n-n_\ell+1}$, $\dots$, $\l_n$ of $K$ and normalize the associated eigenvectors $U \in \R^{n \times (n_s+n_\ell)}$.
		
		\item\label{alg-gap} With the smallest eigenpairs: determine if a
		gap exists, i.e., if there is a $k < n_s$ such that
		\[ \frac{\lambda_{k-1}}{\lambda_{k}}  >
		\mathrm{gf} \times \frac{\lambda_{k}}{\lambda_{k+1}}
		\] If no such $k$ is found, warn user to try a larger $n_s$ and exit. 
		
		\item For each eigenvector \(u_i \), \(i = 1 \text{ to } k\), calculate
		$v^{(e)}:= \half {u_i(m_e)}^T u_i(m_e), \quad e \in \Omega$ for all elements.
		
		
		\item With the largest eigenpairs: for each
		eigenvector $u_i$, calculate $s^{(e)} := \half {u_i(m_e)}^T {T^{(e)}}^T K^{(e)} T^{(e)} u_i(m_e), \quad e \in \Omega, \: i = n-n_l\text{ to }n$.

		\item Normalize $s^{(e)}$ and $v^{(e)}$	s.t. $\max_{e \in \Omega} s^{(e)} = 1$ and $\max_{e \in \Omega} v^{(e)} = 1$ for each eigenvector.
		
		\item Partition $s^{(e)}$ and $v^{(e)}$	into two clusters. 
		
	\end{enumerate}
	\label{alg-modelstab}
\end{algorithm}

\section{Productionising the ML model and presenting results}
\begin{figure} 
	\centering
	\includegraphics[width=0.75\textwidth]{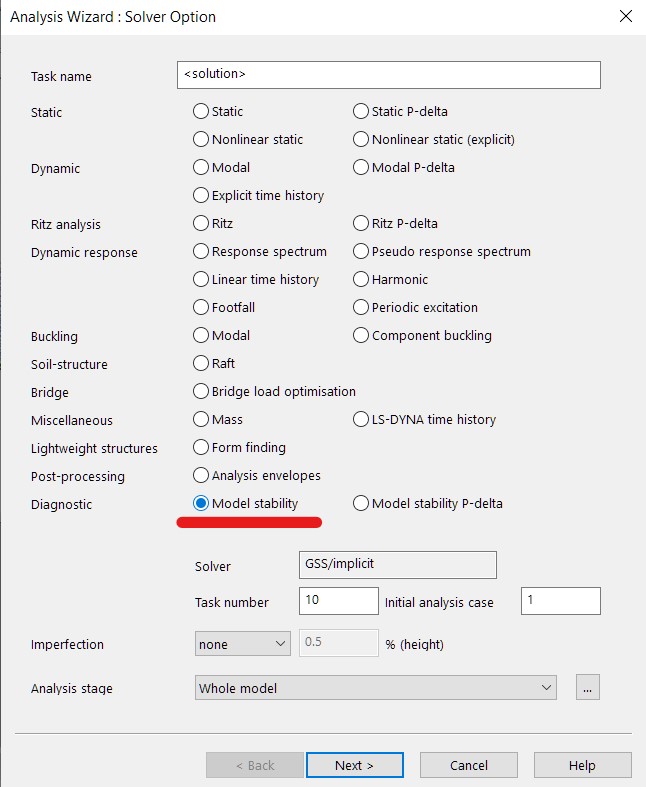} 
	\caption{The algorithm offered as a diagnostic analysis option.}
	\label{fig-analwizard}
\end{figure}

Algorithm \ref{alg-modelstab} uses a stiffness/adjacency matrix and returns a series of clustered (virtual) energy values $v^{(e)}$'s and $s^{(e)}$'s.
As the purpose of the algorithm is to interrogate FE models, a natural place to operationalise this is in the solver software itself, especially since the infrastructure of formulating stiffness and transformation matrices, and manipulating graph adjacency matrices already exists in GSA's solvers.
The crux of the algorithm relies on solving a sparse eigenvalue problem, for which we implemented a bespoke multicore parallel solver. 
The tool was then presented as an ``analysis type" in GSA, similar to linear static analysis or seismic analysis as Figure \ref{fig-analwizard} shows.
We chose this framework as it is familiar and intuitive to engineers.
More precisely Model Stability Analysis is just another analysis that is performed on an FE model, with the difference that it is a diagnostic analysis that checks the stability of the numerical model rather than a structural response analysis.      

\begin{figure} 
	\centering
	\subfloat[3d view]{\includegraphics[width=.5\textwidth]{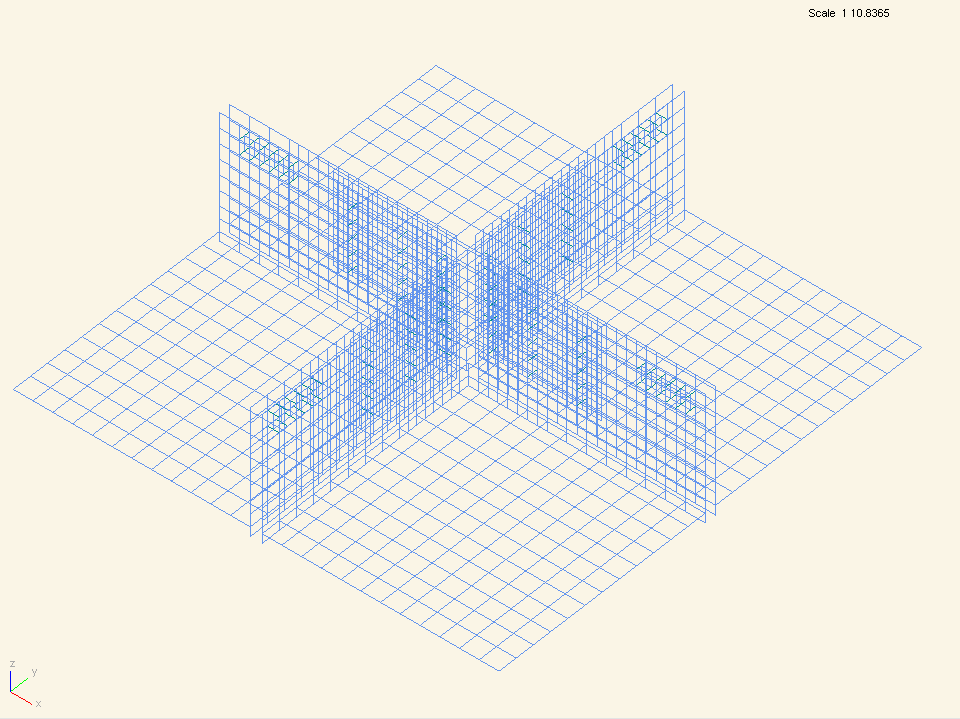}}
	\subfloat[Plan]{\includegraphics[width=.5\textwidth]{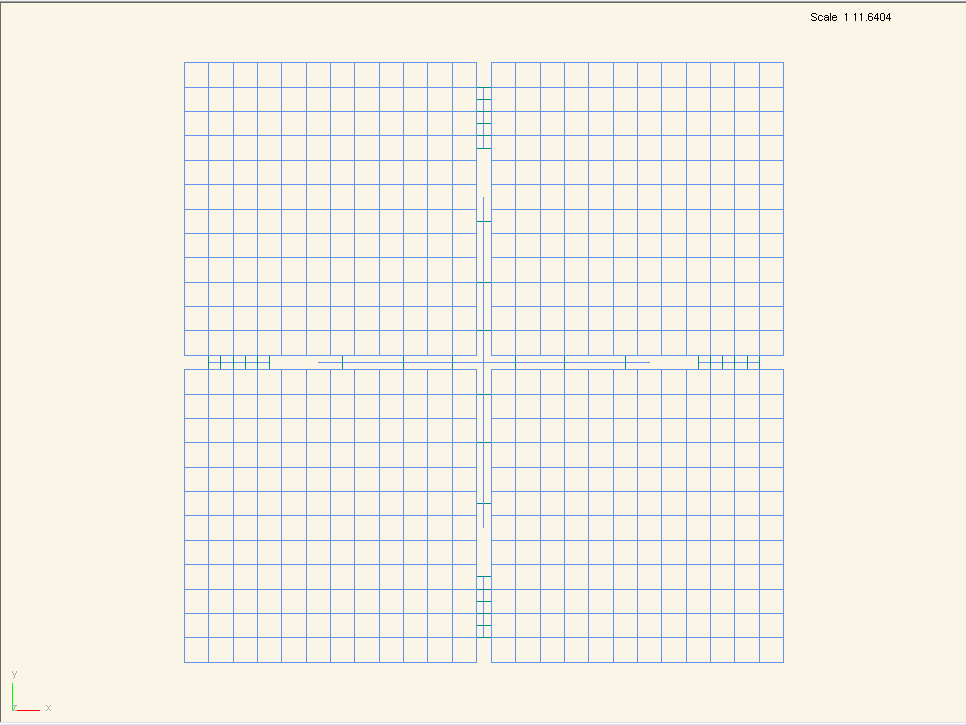}}
	\caption[Connection detail.]{An FE model with beams and quad plate elements containing a modelling error.}
	\label{conn-detail-grview}
\end{figure}

\begin{figure}
	\begin{tabular}{cc}
		\multicolumn{2}{c}{\includegraphics[width=.75\textwidth]{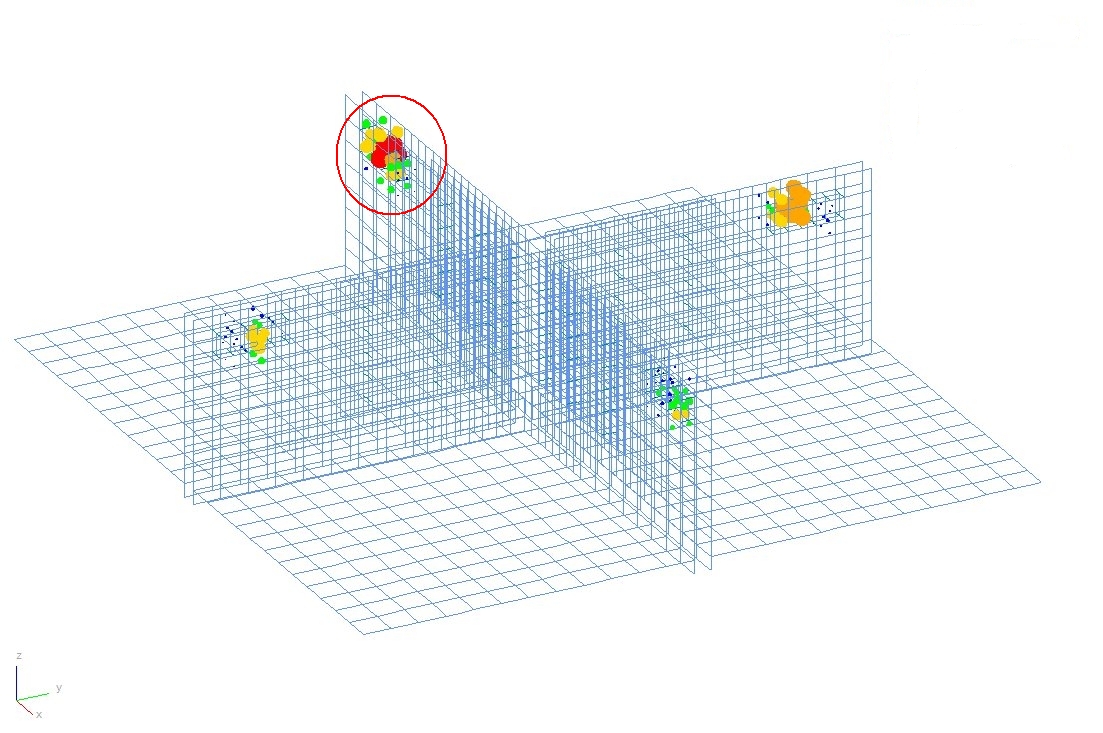}  } \\
		\includegraphics[width=.5\textwidth]{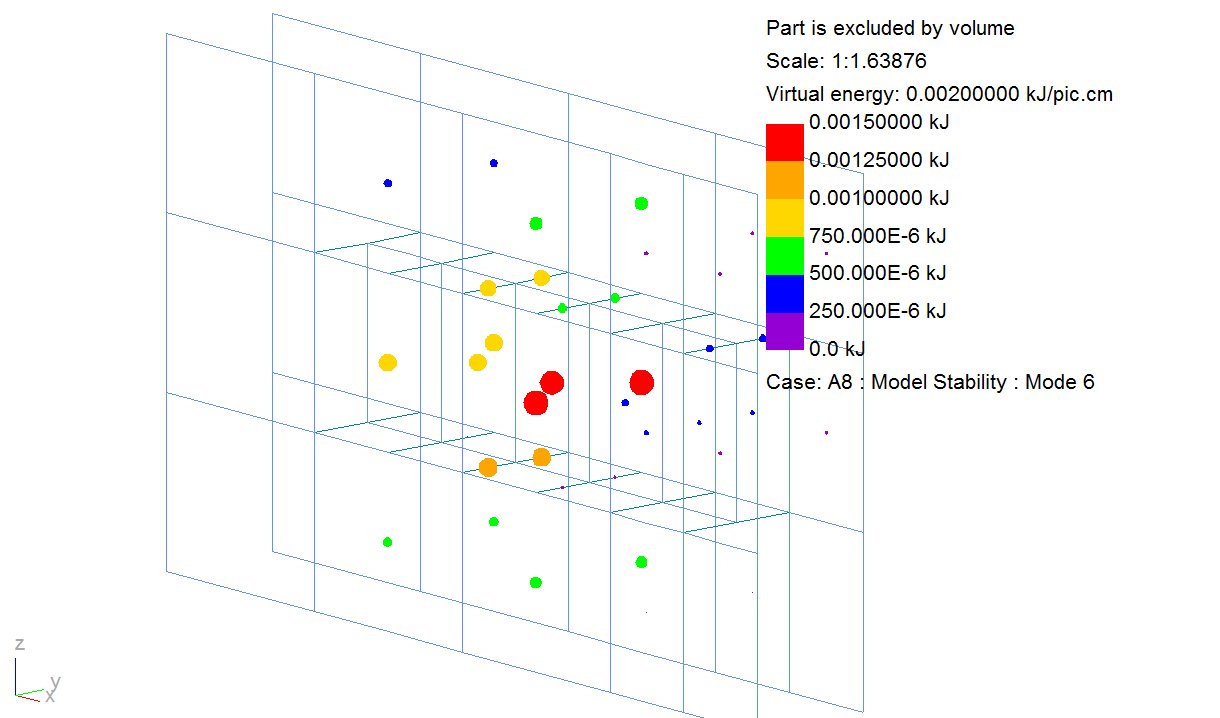} & \includegraphics[width=.5\textwidth]{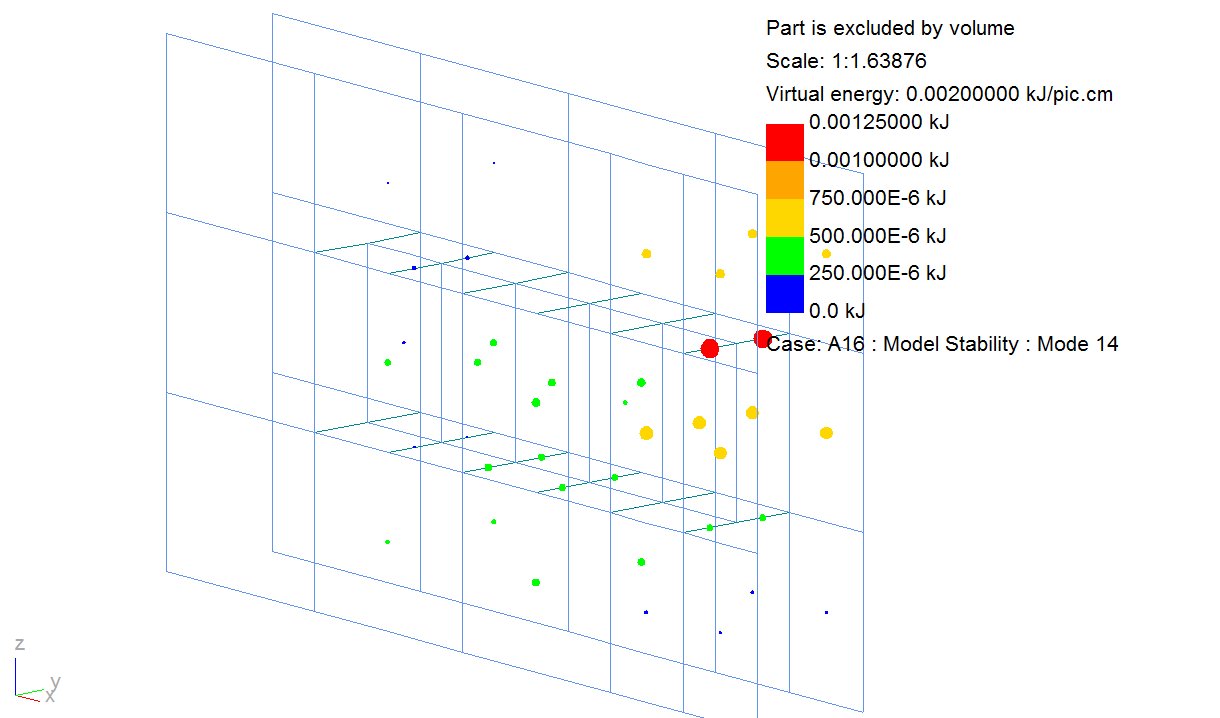}\\
		\includegraphics[width=.5\textwidth]{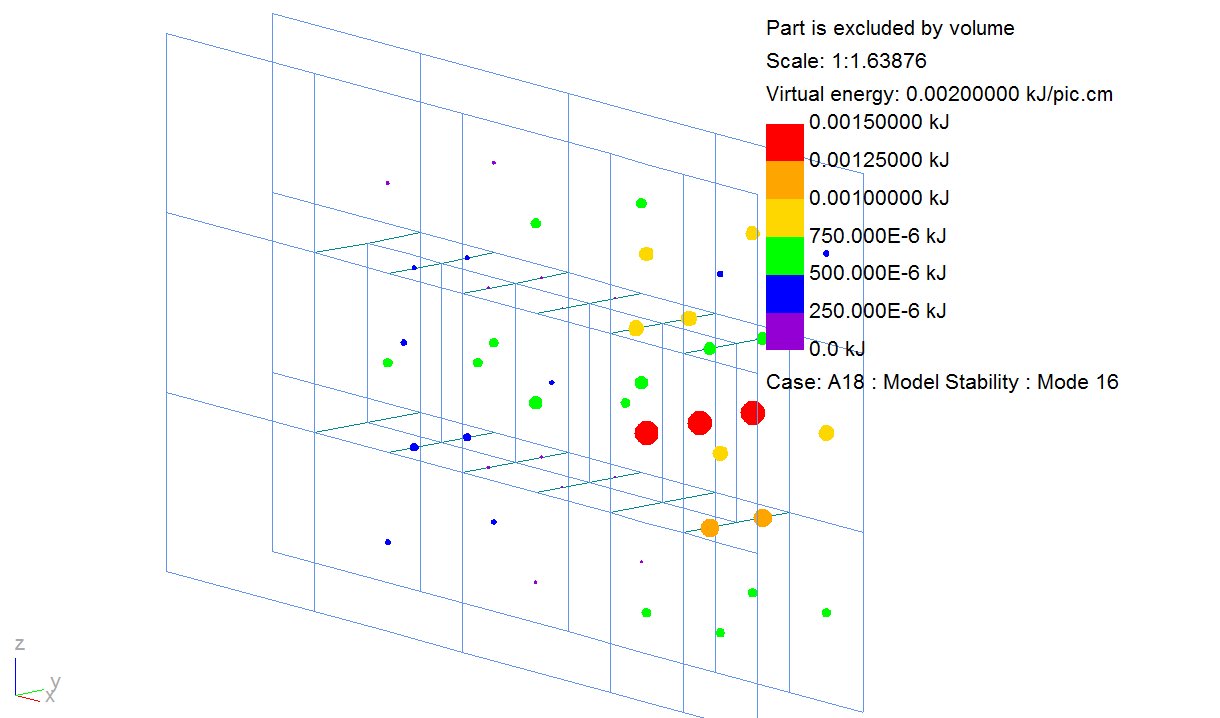} & \includegraphics[width=.5\textwidth]{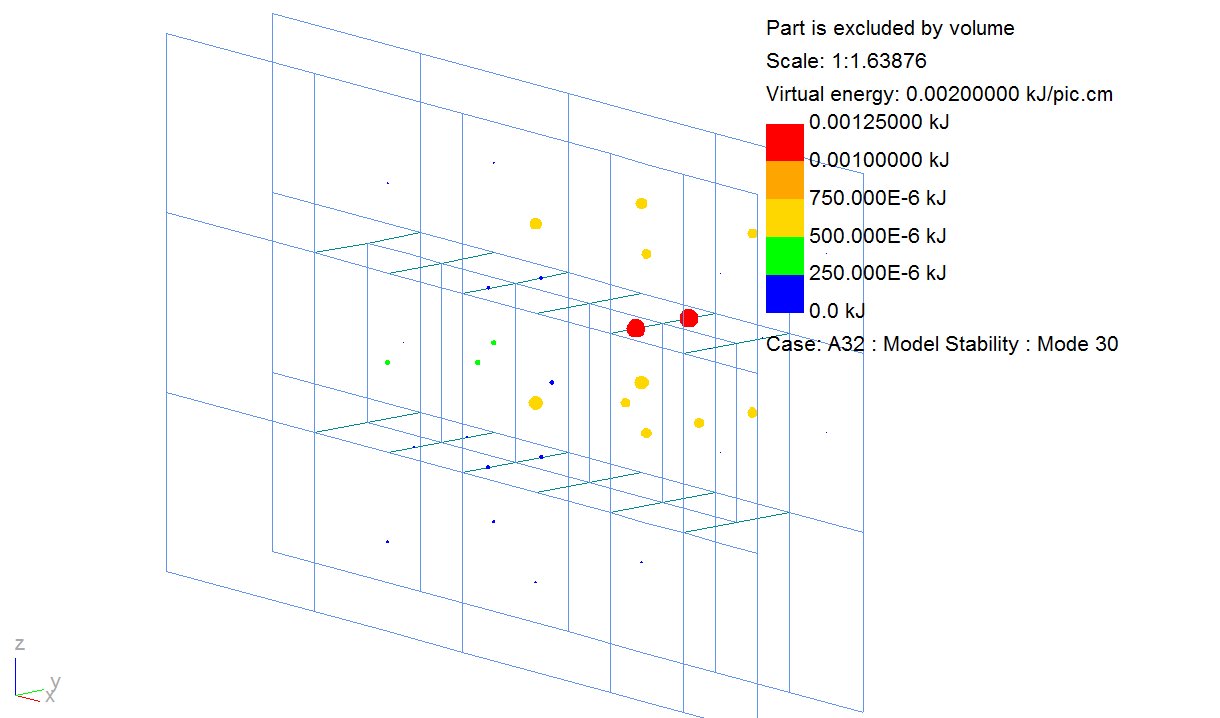} \\
	\end{tabular}
	\caption[Plots for virtual energies for the connection model]{Plots for virtual energies for the FE model. Clockwise from top: energies associated with eigenvector 1, eigenvector 14, eigenvector 30, eigenvector 16 and eigenvector 6. Figures for eigenpairs 6, 14, 16 and 30 are magnified views of the encircled part in eigenvector 1. These values are clustered by mapping out-of-cluster values to 0.}
	\label{fig-cs3-ev-contours}
\end{figure}

\begin{figure}
	\centering
	\includegraphics[width=.75\textwidth]{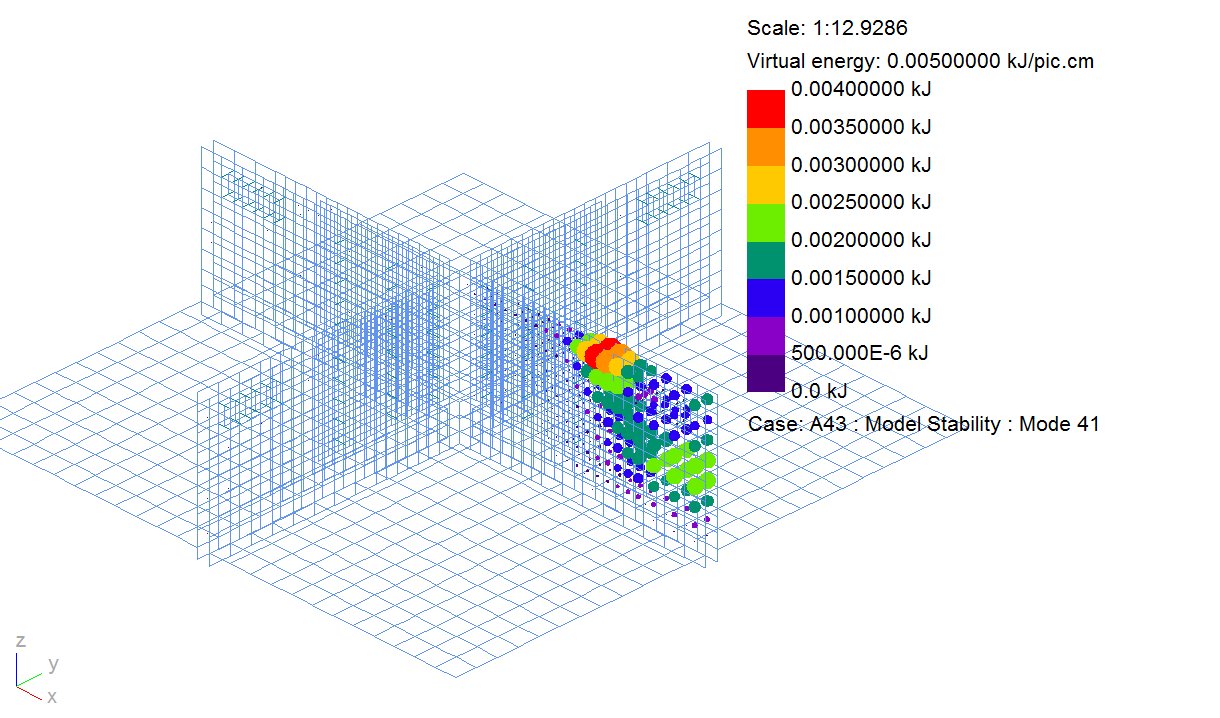}
	\caption[Virtual energy plots for the connection model.]{Contour plots for virtual energies for eigenvector corresponding to the 41st smallest eigenvalue. This eigenvalue is outside the spectral gap, hence does not exhibit clustering.}
	\label{fig-cs3-ev41-contour}
\end{figure}

There are two outputs from the algorithm -- the clusters themselves and the eigenvalues (for assessing the spectral gap).
Clusters are visualized as plots overlaid on the FE model, similar to other analysis results such displacements or stresses. 
The energy values $v^{(e)}$ and $s^{(e)}$'s are normalized such that the largest over all elements is 1 Joule. 
The energies are formulated such that only elements associated with weakly connected dofs have large nonzero energies and all other elements have 0 values.
We then size the radius of each plotted circle using the energy of dof. 
Hence a cluster is naturally visualized when plotting energies overlaid on the FE geometry.
As an example these clusters are shown in Figure \ref{fig-cs3-ev-contours}, where the FE model (Fig. \ref{conn-detail-grview}) had a modelling error whereby incompatible element types shared common nodes. 
The spectral gap for this problem was 40, i.e., the first 40 eigenvectors contained clusters of dofs that caused the ill-conditioning. 
The 41st eigenvector (Figure \ref{fig-cs3-ev41-contour}) does not have pronounced clusters as the first 40.
Further examples are presented in \cite{kahehiti14}.

\section{Summary}
Model Stability Analysis uses a novel spectral clustering variant to allow engineers debug their FE models in GSA.
Over the years since it was first incorporated it has become a useful tool that allows practitioners to gain confidence in their analysis results especially as models increase in size and complexity.

\bibliographystyle{plain}
\bibliography{strings,rkannan}

\end{document}